\documentclass[
    aps,
    prb,
    twoside,
    twocolumn,
    10pt,
    floatfix,
    showpacs,
    citeautoscript,
]{revtex4-2}

\usepackage[T1]{fontenc}
\usepackage{glossaries}
\usepackage{graphicx}
\usepackage[version=4]{mhchem}
\usepackage{siunitx}
\usepackage[colorlinks=true]{hyperref}

\newcommand \Title{
    Impact of Surface Adsorbates and Dimensionality on Templating of Halide Perovskites
}

\hypersetup{
    pdfauthor={E. Fransson, Julia Wiktor, and P. Erhart},
    pdftitle={\Title},
    colorlinks=true,
    linkcolor=black,
    citecolor=black,
    filecolor=black,
    urlcolor=black,
}

\graphicspath{{./figures/}}

\setacronymstyle{long-short}
\newacronym{dft}{DFT}{density functional theory}
\newacronym{hp}{HP}{halide perovskite}
\newacronym{md}{MD}{molecular dynamics}
\newacronym{mlp}{MLP}{machine learning potential}
\newacronym{nep}{NEP}{neuroevolution potential}
\newacronym{ma}{MA}{methylammonium \ce{CH3NH3}}
\newacronym{ba}{BA}{butylammonium \ce{CH3(CH2)3NH3}}
\newacronym{pea}{PEA}{phenylethylammonium \ce{C6H5(CH2)2NH3}}
\newacronym{pma}{PMA}{phenylmethylammonium \ce{C6H5(CH2)NH3}}

\newcommand{\gpumd}{\textsc{gpumd}}
\newcommand{\ovito}{\textsc{ovito}}

\DeclareSIUnit\angstrom{\text{\AA}}
\DeclareSIUnit\atom{\text{atom}}

\newcommand{\addchalmers}{Department of Physics, Chalmers University of Technology, SE-41296, Gothenburg, Sweden}

\begin{document}

\author{Erik Fransson}
\author{Julia Wiktor}
\author{Paul Erhart}
\affiliation{\addchalmers}
\email{erhart@chalmers.se}


\title{\Title}

\begin{abstract}
Two-dimensional (2D) halide perovskites (HPs) are promising materials for various optoelectronic applications, yet a comprehensive understanding of their dynamics is still elusive.
Here, we offer insight into the dynamics of prototypical 2D HPs based on \ce{MAPbI3} as a function of linker molecule and the number of perovskite layers using atomic scale simulations.
We show that the layers closest to the linker undergo transitions that are distinct from those of the interior layers.
These transitions can take place anywhere between a few tens of \unit{\kelvin} below to more than \qty{100}{\kelvin} above the cubic-tetragonal transition of bulk \ce{MAPbI3}.
In combination with the thickness of the perovskite layer this enables one to template phase transitions and tune the dynamics over a wide temperature range.
Our results thereby reveal the details of an important and generalizable design mechanism for tuning the properties of these materials.
\end{abstract}

\maketitle

\begin{figure}[tbh]
\centering
\includegraphics{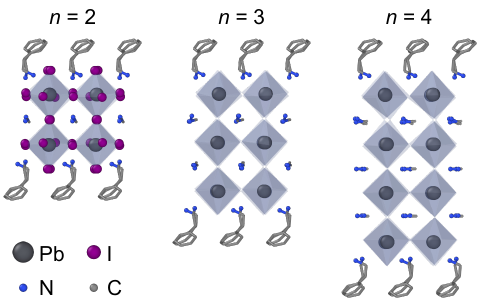}
\caption{
    PEA-based 2D \gls{hp} phases with the composition \ce{PEA2MA_{n-1}Pb_nI_{3n+1}} for $n=2$, 3 and 4 layers.
    For $n>2$ hydrogen and iodine atoms are omitted for clarity.
    The structures were rendered using \ovito{} \cite{Stukowski2010}.
}
\label{fig:structure_visualization}
\end{figure}

\Glspl{hp} are a promising class of materials for various applications, including, e.g., high-efficiency solar cells, \cite{kojima2009organometal, kim2012lead, hodes2013perovskite} lasers \cite{lei2021metal} and light emitting diodes \cite{van2018recent}.
The most studied so far are the regular three-dimensional \glspl{hp} with the formula \ce{AMX3}, where A is an organic or inorganic cation, M is a metal cation, such as Pb or Sn, and X is a halogen.
One of the drawbacks of these compounds is that they often exhibit relatively low stability.
In recent years, so-called two-dimensional (2D) \glspl{hp} (also referred to as layered, quasi-2D or Ruddlesden-Popper phases) \cite{AkkMan20} have, however, gained significant attention \cite{stoumpos2016ruddlesden, cao20152d,tsai2016high, grancini2019dimensional}.
These materials are composed of inorganic perovskite layers stacked on top of each other and separated by organic cations that act as spacers (\autoref{fig:structure_visualization}) \cite{MitFeiHar94, Stoumpos2013b, Mao2018, Ying2021, Liu20023, AkritiPark2024}.
They have been shown to exhibit improved stability \cite{Smith2014, Etgar2018, Hsinhan2018,liu2018tunable, leveillee2019tuning, mahata2021suppression, mosconi2022intermolecular,park2023thickness, Triggs2024} and distinct quantum and dielectric confinement effects \cite{even2014understanding,traore2018composite,KatMerEve19}, which modulate their excitonic properties \cite{DykDuiMau21, Shao2022, ThoDykPek24}, differentiating them from their 3D counterparts.
In combination with their tunability \cite{mahata2019modulating, MihalyiKoch2023}, this makes 2D \glspl{hp} highly attractive for various optoelectronic applications \cite{Etgar2018, fu2019tailoring, chen20182d, qian20192d, MihalyiKoch2023, Shichen2024}.

The properties of 2D \glspl{hp} sensitively depend on the number and type of inorganic layers and the organic cations that connect them \cite{grancini2019dimensional, Mao2018, li20212d, Jahanbakhshi2021, Kingsford2023}.
The inorganic layers are responsible for the electronic structure \cite{Biega2023, Ziegler2022, krach2023emergence} and mechanical properties of the material, while the organic cations affect the interlayer spacing as well as the overall stability and structure.
Therefore, understanding the interplay of inorganic layers and organic cations is crucial for designing efficient and stable optoelectronic devices based on these materials.
This is evident in the so-called ``templating'' approach \cite{mitzi2000organic, mitzi2001templating, xu2003sni42, du2017two, milic2021layered}.
This strategy relies on the fact that the organic linkers can significantly affect the phase of the inorganic framework beyond the surface layer, which can be used to improve the stability of the desired 3D perovskite phases.
To be able to fully exploit the potential of this approach, it is, however, necessary to understand the precise mechanisms by which organic cations influence the inorganic framework.

Here, we offer comprehensive insight into how phase transitions and dynamics in 2D \glspl{hp} can be steered through the choice of the organic linker molecule and the dimensionality of the material.
This is accomplished through atomic scale simulations based on accurate and efficient \glspl{mlp} trained against \gls{dft} calculations.
We first focus on the prototypical combination of the linker molecule \gls{pea} with \ce{MAPbI3} and identify a transition from a high-temperature structure without global octahedral tilting to a lower temperature structure with a global out-of-phase octahedral tilting pattern.
The perovskite layer in direct contact with the \ce{PEA} molecules (referred to as ``surface layer'' below) undergoes a transition already between 450 and \qty{470}{\kelvin}, while the transition in the interior of the perovskite slab occurs at a temperature that is at least \qty{50}{\kelvin} lower.
The combination of these two processes yields a rather broad overall transition, which approaches the transition temperature of bulk \ce{MAPbI3} only for relatively thick inorganic layers comprising at least 30 or more perovskite layers.
To generalize the effect of the linker molecule on the local phase transitions, we then extend the analysis to additional molecules, including \gls{pma}, \gls{ba} and \gls{ma}.
We find that for bulkier molecules like \gls{pea} and \gls{pma}, the surface layer transitions significantly above the bulk \ce{MAPbI3} transition, while with the smallest molecule, \ce{MA}, this transition occurs  at a lower temperature.
Our results thereby provide an atomic scale understanding of how linker and dimensionality can be used to template phase behavior and dynamics in 2D \glspl{hp}.
Since octahedral tilting is intimately tied to the electronic structure \cite{filip2014steric, wiktor2017predictive, zhao2020polymorphous, cannelli2022atomic}, our results reveal the details of an important and generalizable design mechanism for tuning the optoelectronic properties of 2D \glspl{hp}.

\begin{figure}[thb]
\centering
\includegraphics{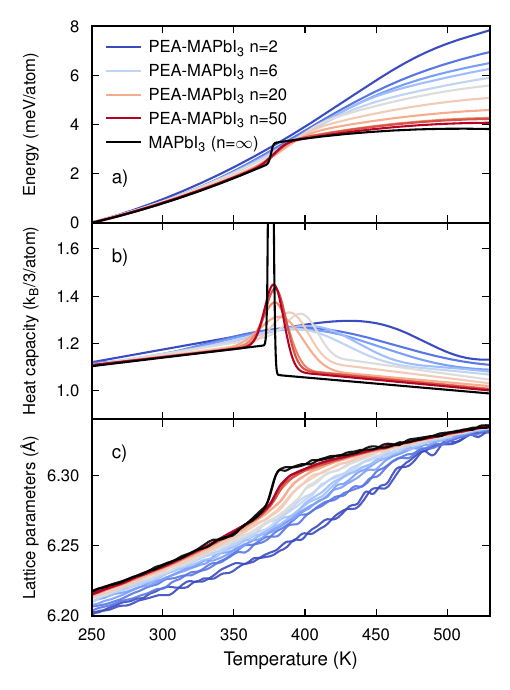}
\caption{
    Thermodynamic observables as a function of temperature from cooling simulations.
    (a) Potential energy (with $1.5k_\text{B}T$ and arbitrary reference energy subtracted) for the a series of 2D \glspl{hp} with composition \ce{PEA2MA_{n-1}Pb_nI_{3n+1}}, which yields \ce{MAPbI3} in the bulk limit ($n\rightarrow\infty$).
    (b) Heat capacity of the system obtained as $C_p=dE/dT$.
    (c) In-plane lattice parameters.
    For \ce{MAPbI3} this corresponds to the $a$ and $b$ lattice parameters and the tilting in the $a^0a^0c^-$ phase occurs around the $z$-axis.
    The potential energy (and heat capacity) shown here are represented by fits to the raw data show in \autoref{sfig:energy_fits}.
}
\label{fig:thermo_data}
\end{figure}

\textbf{Thermodynamic properties}.
We consider a series of 2D \glspl{hp} assembled from inorganic \ce{PbI6} octahedral units with \gls{ma} counterions and \gls{pea} linker molecules with the chemical formula \ce{PEA2MA_{n-1}Pb_nI_{3n+1}}, where $n$ is the number of \emph{perovskite} layers in each inorganic layer (\autoref{fig:structure_visualization}).
In the bulk limit ($n\rightarrow\infty$) one obtains \ce{MAPbI3}, which is one of the most widely investigated 3D \glspl{hp}.
We only consider systems with $n\geq 2$ since in the single perovskite layer limit ($n=1$) we do not observe an untilted inorganic layer even at \qty{600}{\kelvin}.

First, we analyze the potential energy, the heat capacity and the lattice parameters during cooling simulations (\autoref{fig:thermo_data}).
The potential energy of \ce{MAPbI3} shows a small but clear step at \qty{370}{\kelvin}, corresponding to the latent heat associated with its first-order transition from a cubic $a^0a^0a^0$ phase to a tetragonal $a^0a^0c^-$ phase  (\autoref{fig:thermo_data}a) \cite{FraRahWik23}.
This gives rise to a sharp peak in the heat capacity at the transition temperature (\autoref{fig:thermo_data}b).
Additionally, the transition can be seen as a clear change in the two in-plane lattice parameters (tilting is around the out-of-plane axis;  \autoref{fig:thermo_data}c) and even the out-of-plane lattice parameter (\autoref{sfig:heating_cooling}).
The simulations yield a transition temperature for \ce{MAPbI3} of \qty{370}{\kelvin}, which is approximately \qty{40}{\kelvin} higher than the experimental value of about \qty{330}{\kelvin} \cite{WhiHerGui16, Stoumpos2013b}.

Comparable transitions are observed in the two-dimensional \glspl{hp}.
For smaller numbers of inorganic layers, $n$, the transition is more gradual and occurs at higher temperatures, but it becomes more pronounced as $n$ increases, converging toward the behavior observed in \ce{MAPbI3} as $n$ increases.
This shows that the nature of the phase transition evolves from a continuous to a first-order transition.

\begin{figure}[hbt!]
\centering
\includegraphics{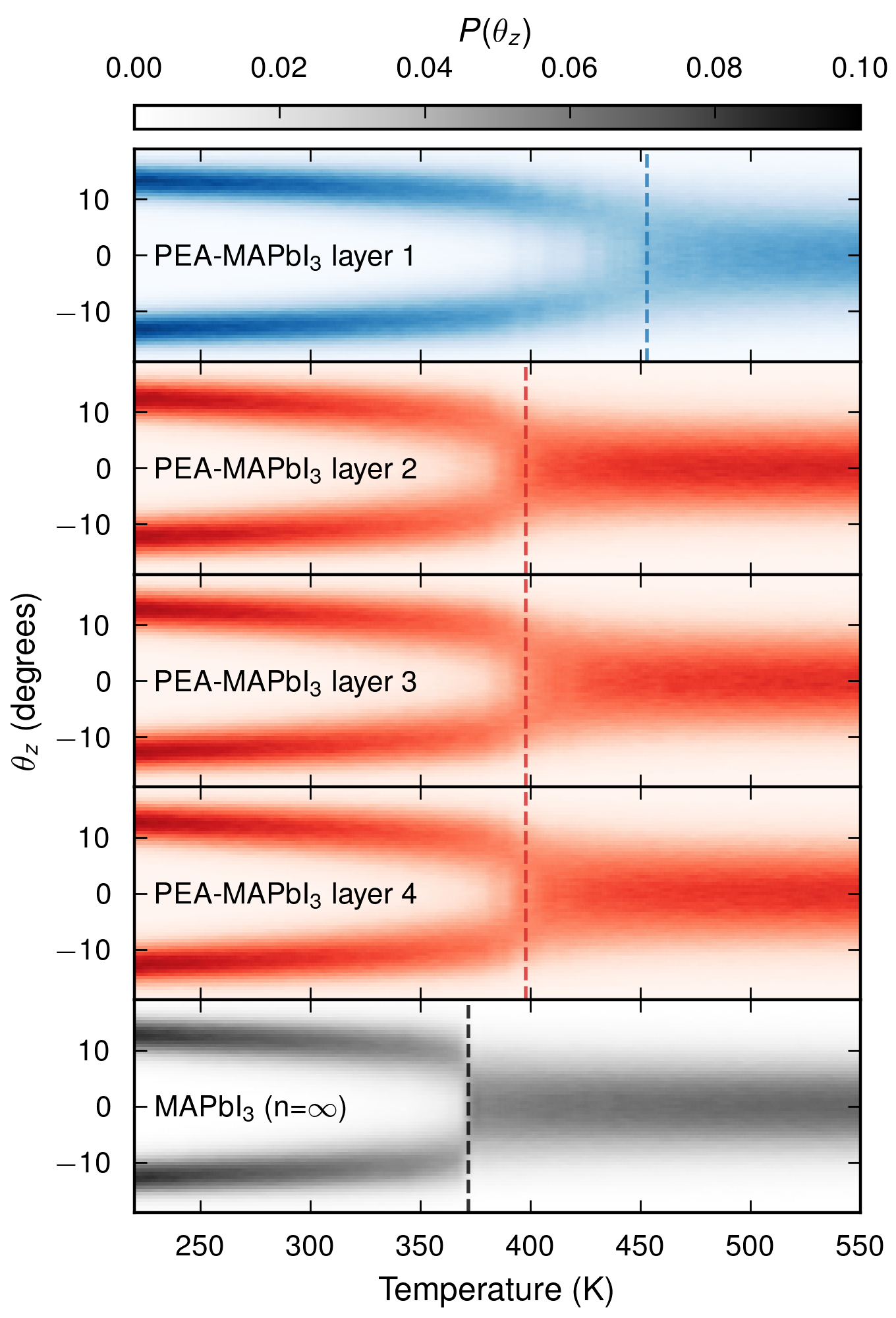}
\caption{
    Distribution over the octahedral tilt angles $P(\theta_z)$ as a function of temperatures for a 2D \gls{hp} \ce{PEA2MA_{n-1}Pb_nI_{3n+1}} with $n=8$ as well as the corresponding 3D \gls{hp} (\ce{MAPbI3}).
    For the 2D \gls{hp} the tilt angle distribution is decomposed by perovskite layer, where layer 1 refers to the perovskite layer closest to the organic linker molecule.
}
\label{fig:tilt_angles_heatmap}
\end{figure}

\begin{figure*}[tb]
\centering
\includegraphics{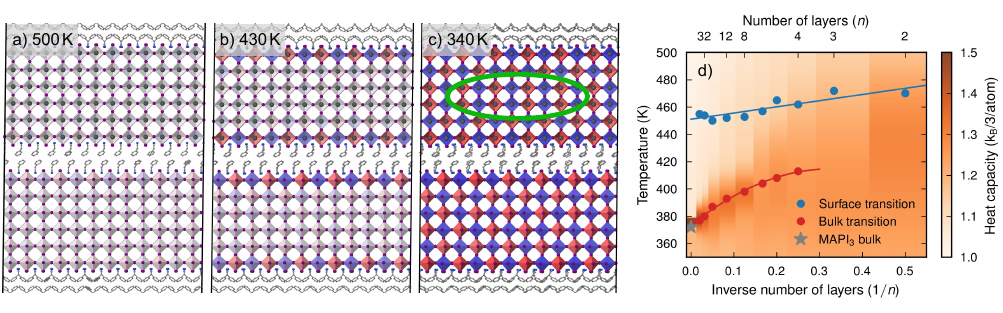}
\caption{
    Time-averaged snapshots from the cooling simulations for the 2D \gls{hp} \ce{PEA2MA_{n-1}Pb_nI_{3n+1}} with $n=6$ at (a) \qty{500}{\kelvin}, (b) \qty{430}{\kelvin} and (c) \qty{330}{\kelvin} visualized using \ovito{} \cite{Stukowski2010}.
    Here, hydrogen atoms as well as the \gls{ma} counterions inside the perovskite layers are omitted for clarity.
    The color coding of the octahedra indicates the rotation angle around the $z$-axis, $\theta_z$, with red and blue indicating negative and positive tilting (ranging from \num{-20} to \qty{+20}{\degree}), respectively, while gray implies tilt angles close to zero.
    For \qty{330}{\kelvin} a stacking fault (anti-phase boundary) is formed as highlighted by the green ellipsoid.
    (d) Transition temperatures as a function of number of layers $n$ with the heat capacity (\autoref{fig:thermo_data}) shown as a heatmap.
}
\label{fig:phase_diagram}
\end{figure*}

\textbf{Octahedral tilting.}
To obtain a more detailed understanding of the transitions we compute the distribution over octahedral tilt angles $P(\theta)$ along the cooling simulations using \ovito{} \cite{Stukowski2010} as done in Ref.~\citenum{WikFraKub2023} (\autoref{fig:tilt_angles_heatmap}).
The tilt angle distribution for a given temperature is averaged over a few snapshots corresponding to a temperature window  of about \qty{1}{\kelvin} in order to improve the statistics.

For bulk \ce{MAPbI3} one observes a sharp transition at \qty{370}{\kelvin} from a single Gaussian peak centered around zero corresponding to a cubic phase ($a^0a^0a^0$) to a symmetric bimodal distribution indicating the transition to a structure with out-of-phase tilting ($a^0a^0c^-$; \autoref{fig:tilt_angles_heatmap}, bottom panel).

\begin{figure*}[tb]
\centering
\includegraphics{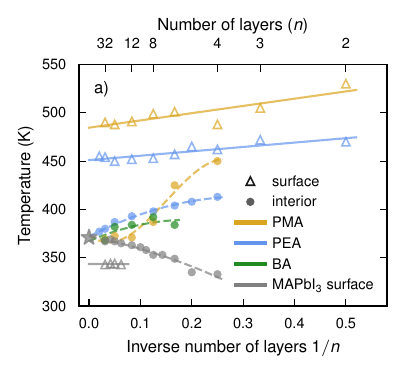}
\includegraphics{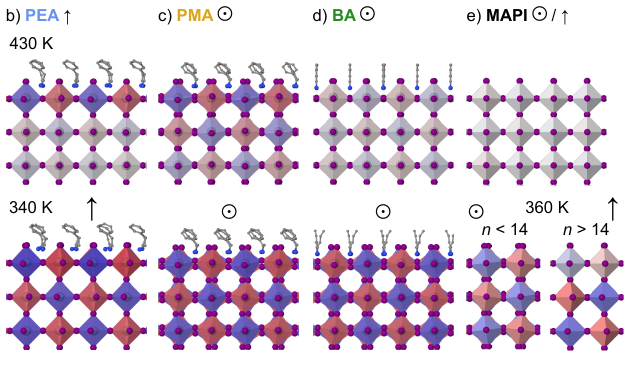}
\caption{
    (a) Transition temperatures as a function of the number of layers $n$ for (b) \gls{pea}, (c) \gls{pma} and (d) \gls{ba}-based 2D \glspl{hp} as well as (e) \ce{MAPbI3} surfaces.
    Triangles and circles indicate the transition temperatures for the surface layer and the interior layers, respectively.
    The star indicates the cubic-tetragonal phase transition temperature for bulk \ce{MAPbI3}.
    (b--e) Average atomic configurations at \qty{430}{\kelvin} (top) and \qty{340}{\kelvin}/\qty{360}{\kelvin} (bottom).
    Red and blue octahedra indicate negative and positive tilt angles (ranging from \num{-20} to \qty{+20}{\degree}), respectively, whereas gray implies tilt angles close to zero.
    Arrows indicate the tilt axis, which is out-of-plane for \gls{pea} and \ce{MAPbI3} surfaces with less than 14 layers, and in-plane for the other systems.
    Lines in (a) serve as a guide to the eye.
    }
    \label{fig:phase_diagram_all}
\end{figure*}

For the 2D \glspl{hp} we can resolve the tilt angle distribution for each symmetrically distinct perovskite layer throughout the structure.
This analysis reveals that the perovskite layer that is in direct contact with the \gls{pea} linker molecules (the ``surface layers'') undergoes a transition to a tilted structure that for, e.g., $n=8$ occurs at around \qty{450}{\kelvin} (\autoref{fig:tilt_angles_heatmap}; top panel).
In contrast, the interior perovskite layers undergo a transition at a much lower temperature, i.e., closer to the bulk \ce{MAPbI3} transition temperature, e.g., at around \qty{400}{\kelvin} for $n=8$, .
It is worth noting that the transition in the surface layers has almost no impact on the tilting in the neighboring layer (layer 2 in \autoref{fig:tilt_angles_heatmap}; also compare \autoref{fig:phase_diagram}b and \autoref{fig:phase_diagram_all}b).
We attribute this behavior to the octahedra rotating around the $z$-axis, leading to a weak correlation between neighboring octahedra in the $z$-direction \cite{WikFraKub2023, Baldwin2023}.

At high temperatures, for which no global tilting pattern occurs, the tilt angle distributions are unimodal and well described by Gaussians with zero mean.
The width of the distribution is, however, wider for the surface layers compared to the rest of the layers indicating a softer free energy landscape.
Furthermore, at low temperature, for which all octahedra exhibit a tilt, the surface layers show slightly larger tilt (\autoref{sfig:PEA_angle_slice}).
Both of these observations are consistent with the surface layers exhibiting a higher transition temperature.

\textbf{Phase diagram.}
The spatial variation of the evolution of octahedral tilts means that the \gls{pea} based 2D \glspl{hp} internally undergo two transitions that can be observed separately in our simulations.
The first one is associated with the tilting of the octahedra in the \emph{surface} layer, while the second one is related to the tilting of the \emph{interior} layers.
Extending the tilt-angle analysis for $n$ ranging from 2 to 50 allows us to obtain the variation of the two transition temperatures with $n$ (\autoref{fig:phase_diagram}).
(For a brief discussion of the uncertainties in the transitions temperatures please see \autoref{sect:transitions} in the Supplemental Material.)
This shows that the transition in the surface layer depends only weakly on $n$ varying from \qty{470}{\kelvin} ($n=2$) to about \qty{450}{\kelvin} (large-$n$ limit).
The transition in the interior, which can only be identified for $n\geq 4$, exhibits a more pronounced dependence on $n$ starting at about \qty{410}{\kelvin} for $n=4$ and converging to the bulk \ce{MAPbI3} value of \qty{370}{\kelvin} in the large-$n$ limit.

The different structure of the surface layer compared to the interior resembles surface (interface) phases, also referred to as complexions \cite{Cantwell2014, CanFroRup2020}.
This type of surface phases can be understood from a simplified thermodynamic viewpoint using surface and interface free energies $\gamma$ \cite{Johansson2011}.
In this view, the above observation suggests that the effective interface free energy between the cubic phase and the organic linkers $\gamma_\text{cub/PEA}$ is larger than the sum of the interface energy between the tetragonal phase and the organic linkers $\gamma_\text{cub/tet}$, and the tetragonal and cubic phases $\gamma_\text{tet/PEA}$, i.e., $\gamma_\text{cub/PEA} > \gamma_\text{cub/tet} + \gamma_\text{tet/PEA}$.

In our simulations, the tilting of the two surface layers on the opposite sides of the inorganic slab are not correlated with each other at the upper transition temperature and can thus occur by chance in-phase or out-of-phase.
For the out-of-phase tilting pattern ($a^0a^0c^-$) to be commensurate with both surface layers, the latter need to tilt out-of-phase or in-phase with respect to each other for an even and odd number of layers, $n$, respectively.
As a result anti-phase boundaries can be expected to appear with 50\% probability at nucleation time and are commonly observed in our simulations (\autoref{fig:phase_diagram}).
In some cases we observe such defects to anneal out already on the time scale of our simulations.
Under experimental settings one can therefore assume that such defects typically anneal out and are only present in small concentrations.

Lastly, we look at the ordering of the linker molecules.
The two layers of \gls{pea} forming a single organic spacer layer are always rotated \qty{180}{\degree} around the $z$-axis relative to each other (\autoref{fig:structure_visualization}).
In addition, we observe that the different spacer layers can take on arbitrary 90 and \qty{180}{\degree} rotations around the $z$-axis (see e.g., \autoref{fig:phase_diagram}).
This leads to the in-plane lattice parameters being equal (see \autoref{fig:thermo_data}).
Reorientation and rotation of the spacer layers mainly take places during the equilibration part of the simulations, and appear to occur statistically.
The orientation subsequently remains largely unaffected as temperature is reduced.

\textbf{Extension to other systems.}
Now that we have seen how \gls{pea} templates the phase transition in the perovskite layers, it is instructive to extend the analysis to other linker molecules.
To this end, we consider 2D \glspl{hp} based on \gls{pma} and \gls{ba} as well as \gls{ma}-terminated surfaces, specifically \{001\} slabs of \ce{MAPbI3} with \ce{MAI2} termination (\autoref{fig:phase_diagram_all}).

For \textbf{\gls{pma}} the behavior is qualitatively similar to that of \gls{pea} (\autoref{sfig:PMA_angles}), i.e., a transition of the octahedral tilting pattern occurs in the surface layer at a temperature about \qty{100}{\kelvin} higher than in the interior, albeit with a stronger dependence on the number of layers for the interior transition.
Unlike the case of \gls{pea} for which we found tilting with respect to the out-of-plane axis ($z$), with \gls{pma} we obtain tilting around one of the in-plane axes ($x$ or $y$).

For both \gls{pea} and \gls{pma} we observe that the transition temperature for the interior increases notably with decreasing number of layers, approaching the transition temperature for the surface for the thinnest slabs considered here.
This reflects the increasing relative weight of the surface layer compared the rest of the system as $n$ decreases.
Similarly in the limit of large $n$, the interior transition temperature approaches that of bulk \ce{MAPI3}.

We also note that in the case of \gls{pma} we observe almost no anti-phase boundaries.
We suggest this to be due to the stronger octahedral correlation \emph{perpendicular} compared to \emph{along} the rotational axis, as previously reported in bulk \glspl{hp} \cite{WikFraKub2023, Baldwin2023}.
This likely leads to a stronger driving force for the (re)orientation of perovskite layers which is needed to avoid or anneal out anti-phase boundaries.

By contrast, in the case of \textbf{\gls{ba}}, we observe no separation in temperature between the onset of tilting at the surface and the interior.
Rather, there is just one transition that for the smallest $n$ is barely 10 to \qty{20}{\kelvin} higher than the phase transition temperature for bulk \ce{MAPbI3} with a very weak dependence on the number of layers.
Similarly to the case of \gls{pma}, for \gls{ba} the tilting occurs around one of the in-plane axes.
We note that one can observe a secondary transition associated with the motion and ordering of the \gls{ba} linker molecules themselves (\autoref{sfig:BA_second_transtion}).
At higher temperatures the \gls{ba} molecules move much more freely than \gls{pea} and \gls{pma} \cite{Biega2023}, and are on average oriented perfectly perpendicular to the perovskite layers.
Below \qty{300}{\kelvin} this motion is, however, frozen out and the \gls{ba} molecules become significantly stiffer.

For the \textbf{\ce{MAPbI3} surface} we observe two different types of behavior.
For thicker slabs ($n > 14$) the topmost (surface) layer undergoes a transition at a \emph{lower} temperature than the interior region, thus exhibiting the opposite behavior compared to \gls{pea} and \gls{pma}.
On the other hand, for thinner slabs ($n < 14$) the surface transition can no longer be separated from the transition in the interior of the slab.
This can be at least partly explained by the transition temperature for the interior layers decreasing with the number of layers which causes the surface-to-interior ratio to increase.
We also also observes a qualitative difference in the tilt pattern between thicker and thinner slabs as the former exhibit tilting with respect to the out-of-plane axis while for the latter tilting occurs with respect to one of the in-plane axes.
This behavior suggests that the balance between surface and bulk energetics plays a key role here.
While resolving the mechanism is beyond the scope of the present work it is deserving of a more in-depth analysis in future studies.

To summarize our analysis indicates that tilting behavior of the surface layer in 2D \ce{MAPbI3}-based perovskites, i.e., the softness of the rotational energy landscape of the octahedra, can be altered and controlled through the choice of the organic linker molecule.
For the bulkier molecules, \gls{pea} and \gls{pma}, we find that the surface layer transitions at a considerably \emph{higher} temperature than bulk \ce{MAPbI3}, whereas for the smallest molecule considered here, \ce{MA}, we rather observe the surface transition to occur at a \emph{lower} temperature than in the bulk.
This leads to a transition temperature for the \emph{interior} that decreases and increases with the number of layers for \ce{PEA}/\ce{PMA} and \ce{MA}, respectively.
For \ce{BA} an intermediate behavior is observed, i.e., no separate transition for the surface layer.
These results thus provide guiding principles for how both dimensionality (through the number of layers $n$) and chemistry (through the organic linkers) can be used to \emph{systematically} tune the structural transitions and consequently the inorganic dynamics of the system.
Both of these are directly tied to enhanced electron-phonon coupling, which is at the heart of the outstanding optoelectronic properties of these materials.
The present insight is thereby of immediate interest for designing 2D \gls{hp} materials and devices for specific applications and temperature ranges.

\textbf{Acknowledgments.}
This work was supported by the Swedish Research Council (grant numbers 2020-04935, 2021-05072), the Chalmers Initiative for Advancement of Neutron and Synchrotron Techniques, the Swedish Strategic Research Foundation through a Future Research Leader programme (FFL21-0129) and the Wallenberg Academy Fellow program (J.~W.).
The computations were enabled by resources provided by the National Academic Infrastructure for Supercomputing in Sweden (NAISS) at C3SE, NSC, and PDC partially funded by the Swedish Research Council through grant agreements no. 2022-06725 and no. 2018-05973 as well as the Berzelius resource provided by the Knut and Alice Wallenberg Foundation at NSC.

We thank Göran Wahnström, Rasmus Lavén, Maths Karlsson, Prakriti Kayastha and Lucy Whalley for helpful discussions on 2D perovskites.

\textbf{Computational Methods.}
The PEA-based 2D \glspl{hp} have a composition of \ce{PEA2MA_{n-1}Pb_nI_{3n+1}} where $n$ corresponds to the number of perovskite layers.
Starting from known prototypes for $n=1$ \cite{du2017two, Menahem2021, Zuri2023, Liu20023}, we construct structures with $n>1$ by inserting the required number of perovskite layers (\autoref{fig:structure_visualization}).
These structures are then equilibrated by \gls{md} simulations at \qty{600}{\kelvin} to remove structural bias before the cooling simulations.
This approach is also employed for \gls{pma} and \gls{ba} using the prototype structures from Refs.~\citenum{Papavassiliou1999, du2017two, Menahem2021}.

Energies, forces and virials were obtained for the training structures via \gls{dft} calculations as implemented in the Vienna ab-initio simulation package \cite{KreHaf93, KreFur1996-1, KreFur1996-2} using the projector augmented wave method \cite{Blo94, KreJou99} with a plane wave energy cutoff of \qty{520}{\electronvolt} and the SCAN+VV10 exchange-correlation functional \cite{PenYanPer16}.
The Brillouin zone was sampled with automatically generated $\boldsymbol{k}$-point grids with a maximum spacing of \qty{0.25}{\per\angstrom}.

We constructed a \gls{nep} model using the iterative strategy outlined in Ref.~\citenum{FraWikErh2023} using the \gpumd{} software \cite{FanZenZha21, Fan22, FanWanYin22}.
Training structures included \gls{md} structures at various temperatures up to \qty{600}{\kelvin} for bulk \ce{MAPbI3}, 2D \gls{hp} structures with varying number of perovskite layers and three different organic linkers, \gls{pea}, \gls{pma} and \gls{ba}.
Additionally, prototype (primitive) structures with varying volume were included as well as a few dimer configurations.
The \gls{md} structures were generated via an active learning strategy using earlier \gls{nep} model generations and selected according to their uncertainty, which was estimated from the predictions of an ensemble of models.
The final \gls{nep} was then trained using all available training data.
In total the training set consists of \num{616} structures, corresponding to a total of \num{120000} atoms.
For the final model, the root mean squared errors obtained by cross validation using 10 folds are \qty{10}{\milli\electronvolt\per\atom} for the energies, \qty{150}{\milli\electronvolt\per\angstrom} for the forces and \qty{90}{\milli\electronvolt\per\atom} for the virials (\autoref{sfig:parity_plots}).

All \gls{md} simulations were carried out with \gpumd{} \cite{Fan17, FanWanYin22} with a timestep of \qty{0.5}{\femto\second}.
The cooling simulations were run in the NPT ensemble by first heating the system up from zero to \qty{600}{\kelvin} over \qty{1}{\nano\second}, followed by equilibration at \qty{600}{\kelvin} for \qty{1}{\nano\second}, before finally cooling down to \qty{200}{\kelvin} over \qty{25}{\nano\second}.
Simulations were carried using cells comprising \numproduct{6x6x4} repetitions of the 2D prototype structures which for, e.g., $n=12$ corresponds to about \num{50000} atoms and a cell size of about \qtyproduct{50x50x350}{\angstrom}; see \autoref{sfig:cooling_conv} for convergence testing.

The transitions between different perovskite phases were analyzed using the octahedral tilt angles of the \ce{PbI6} octahedra \cite{WikFraKub2023, Fransson2023, Baldwin2023, Liang2023}.
The tilt angles in the perovskite layers during \gls{md} simulations were obtained using \ovito{} \cite{Stukowski2010} as implemented in Ref.~\citenum{WikFraKub2023}.

\textbf{Data availability.}
The \gls{dft} data and \gls{nep} model generated in this study are publicly available via Zenodo at \url{https://doi.org/10.5281/zenodo.11120638}.

\end{document}